\documentclass[sigconf]{acmart}

\usepackage{amsmath}
\usepackage{amsfonts}
\usepackage{makecell}
\usepackage{bm}
\usepackage{enumitem}
\AtBeginDocument{%
  }

\setcopyright{acmcopyright}

\copyrightyear{2023}
\acmYear{2023}
\setcopyright{rightsretained}
\acmConference[ICAIF '23]{4th ACM International Conference on AI in Finance}{November 27--29, 2023}{Brooklyn, NY, USA}
\acmBooktitle{4th ACM International Conference on AI in Finance (ICAIF '23), November 27--29, 2023, Brooklyn, NY, USA}
\acmDOI{10.1145/3604237.3626913}
\acmISBN{979-8-4007-0240-2/23/11}

\begin{document}

% \title{Extended Chiarella Model with Stochastic Volatility with Applications in Deep Hedging}
% \title{Optimising Deep Hedging with Agent-based Financial Market Simulation}
% \title{Deeper Hedging: A Chiarella-Heston Model with Applications in Deep Hedging}
\title{Deeper Hedging: A New Agent-based Model for Effective Deep Hedging}

\author{Kang Gao}
\email{kang.gao18@imperial.ac.uk}
% \author{123}
\affiliation{
  \institution{Imperial College London}
  \city{London}
  \country{United Kingdom}
}

\author{Stephen Weston}
\email{sweston@deloitte.co.uk}
\affiliation{
  \institution{Deloitte LLP}
  %\streetaddress{1 Th{\o}rv{\"a}ld Circle}
  \city{London}
  \country{United Kingdom}}

\author{Perukrishnen Vytelingum}
\email{krishnen@simudyne.com}
\affiliation{%
  \institution{Simudyne Limited}
  \city{London}
  \country{United Kingdom}
}

\author{Namid R. Stillman}
\email{namid@simudyne.com}
\affiliation{%
  \institution{Simudyne Limited}
  \city{London}
  \country{United Kingdom}
}

\author{Wayne Luk}
\email{w.luk@imperial.ac.uk}
\affiliation{%
  \institution{Imperial College London}
  \city{London}
  \country{United Kingdom}}

\author{Ce Guo}
\email{c.guo@imperial.ac.uk}
\affiliation{%
  \institution{Imperial College London}
  \city{London}
  \country{United Kingdom}}

% \renewcommand{\shortauthors}{Gao et al.}
\renewcommand{\shortauthors}{Kang Gao, Stephen Weston, Perukrishnen Vytelingum, Namid Stillman, Wayne Luk, and Ce Guo}
\newcommand{\ce}[1]{{#1}}
\begin{abstract}
We propose the Chiarella-Heston model, a new agent-based model for improving the effectiveness of deep hedging strategies. This model includes momentum traders, fundamental traders, and volatility traders. The volatility traders participate in the market by innovatively following a Heston-style volatility signal. The proposed model generalises both the extended Chiarella model and the Heston stochastic volatility model, and is calibrated to reproduce as many empirical stylized facts as possible. According to the stylised facts distance metric, the proposed model is able to reproduce more realistic financial time series than three baseline models: the extended Chiarella model, the Heston model, and the Geometric Brownian Motion. The proposed model is further validated by the Generalized Subtracted L-divergence metric. With the proposed Chiarella-Heston model, we generate a training dataset to train a deep hedging agent for optimal hedging strategies under various transaction cost levels. The deep hedging agent employs the Deep Deterministic Policy Gradient algorithm and is trained to maximize profits and minimize risks. Our testing results reveal that the deep hedging agent, trained with data generated by our proposed model, outperforms the baseline in most transaction cost levels. Furthermore, the testing process, which is conducted using empirical data, demonstrates the effective performance of the trained deep hedging agent in a realistic trading environment.

\end{abstract}

\begin{CCSXML}
<ccs2012>
<concept>
<concept_id>10010147.10010341</concept_id>
<concept_desc>Computing methodologies~Modeling and simulation</concept_desc>
<concept_significance>500</concept_significance>
</concept>
<concept>
<concept_id>10003752.10010070.10010071.10010261</concept_id>
<concept_desc>Theory of computation~Reinforcement learning</concept_desc>
<concept_significance>500</concept_significance>
</concept>
</ccs2012>
\end{CCSXML}

\ccsdesc[500]{Theory of computation~Reinforcement learning}
\ccsdesc[500]{Computing methodologies~Modeling and simulation}

\keywords{Agent-based Modelling, Stochastic Volatility Models, Deep Reinforcement Learning, Deep Hedging\newline}



\maketitle

\section{Introduction}
In recent years, the quantitative finance industry has emerged as a leading field in the application of data science. With a rapidly growing volume of data and increased computing power available, complex and time-consuming algorithms can be applied to a range of practical applications. In terms of hedging derivatives, financial institutions have begun to explore deep learning algorithms as a replacement for traditional sensitivity-based approaches. In an idealized world with a complete market, perfect hedging is accessible based on the Black-Scholes model \cite{blackscholes}. However, the real market always involves various frictions, such as transaction costs and slippage, making hedging optimization a much more challenging problem. Since the analytic formulas are \ce{not} available in such a market, it may be necessary to hedge derivatives based on \ce{non-closed-form computational procedures or human expertise.}

% Leveraging the powerful representational capabilities of neural networks and advanced optimization algorithms, training a neural network is likely to achieve an optimal hedge. 
% Notably, the experiments \ce{in} \cite{deephedging} and \cite{imaki2021no} demonstrate that Deep Hedging algorithms exhibit high feasibility and scalability when applied to hedging under transaction costs. The deep hedging method represents an alternative to traditional hedging methods that often rely on pre-defined \ce{hard-coded} hedging strategies. By utilizing the deep hedging method, institutions can eliminate the need for a priori assumptions and instead rely entirely on data-driven strategies. 

Deep hedging \cite{deephedging} presents an innovative approach to enhance hedging operations. Within this framework, a neural network is employed to hedge derivatives effectively by minimizing a suitable risk measure. Deep hedging utilises deep reinforcement learning agents to derive hedging strategies. As a data-driven methodology, deep hedging requires a large amount of data input for training the deep reinforcement learning agent. The majority of the literature has mainly been training and testing deep reinforcement learning agents using synthetic data. For example, \cite{deephedging} uses a dataset generated by Heston stochastic volatility model, while \cite{cao2021deep} utilises Geometric Brownian Motion and SABR stochastic volatility model. The underlying assumption is that an agent trained by synthetic data can be transferred to the real environment successfully. This assumption requires that the synthetic data used for training the agents are able to closely mimic real financial market dynamics. However, this is rarely true in practice. For example, the commonly used Geometric Brownian Motion assumes fixed volatility of the underlying stock, while in reality the volatility constantly changes and exhibits a volatility clustering stylised fact. To enhance the performance of deep hedging agents in real financial markets, advanced financial market simulators are essential to generate a realistic training dataset.

Surprisingly, the literature on deep hedging barely uses synthetic data generated by agent-based models. Agent-based models (ABMs) are computational models used to simulate the actions and interactions of autonomous agents in a network or environment to analyze complex systems. In the context of financial markets, agents could represent traders, investors, or financial institutions, each with their own behaviours, strategies, and decision-making processes. With huge potential academic and industrial value, agent-based financial market simulation has gained extensive research attention in recent years. Agent-based models are particularly useful for generating synthetic financial market data that closely resemble real market dynamics. Empirical financial time series data exhibit various stylised facts \cite{cont2001empirical}. Calibrated to reproduce these stylized facts, agent-based models present a clear advantage over traditional equilibrium-based mathematical models when it comes to simulating financial markets.

The primary challenge that we address includes designing an agent-based model, calibrating the model to generate realistic financial time series data, as well as utilising the proposed agent-based model to enhance the performance of the deep hedging agent in a practical environment. We address this challenge by proposing the Chiarella-Heston model, a novel agent-based model. The model is calibrated to reproduce stylised facts exhibited by empirical data. Realistic financial time series data are generated by the model and used for training a deep hedging agent. To demonstrate the practical value of our methodology, we test the trained deep hedging agent on empirical data. The main \textbf{contributions} of our work are the following:
\begin{itemize}[topsep=10pt, partopsep=5pt]
    \item An innovative agent-based model called Chiarella-Heston model is proposed, comprising momentum traders, fundamental traders, and volatility traders. The model is a generalisation of both the extended Chiarella model in \cite{MAJEWSKI2020103791} and the Heston stochastic volatility model in \cite{heston1993closed}. The model is calibrated to reproduce empirical stylised facts including fat tails of returns and autocorrelation patterns of returns. Using popular validation metrics we show that our agent-based model is capable of producing more realistic financial time series than baseline models including the extended Chiarella model and the Heston stochastic volatility model. \newline
    \item With the calibrated agent-based financial market simulator, a training dataset is generated to train the deep hedging agent for optimal hedging strategies under different transaction cost levels. Empirical results show that at most transaction cost levels, the deep hedging agent trained by data generated by the proposed Chiarella-Heston model outperforms the baseline, where the deep hedging agent is trained by data generated by Geometric Brownian Motion. The deep hedging agent utilises the Deep Deterministic Policy Gradient algorithm \cite{lillicrap2015continuous} and is trained towards maximising profits and minimising risks. In addition, our testing process is carried out using empirical data, showing that the trained deep hedging agent is able to perform well in realistic trading environments. 
\end{itemize}

\section{Related work}

\subsection{Agent-based Financial Market Simulation}

An agent-based model (ABM) is a computational simulation that operates based on the individual decisions made by programmed agents \cite{todd2016agent}. ABMs find widespread application in simulating financial markets. In agent-based simulated financial markets, each agent aims to process the vast amounts of time series information generated during the market simulation and translate it into trading decisions \cite{lebaron2001builder}. The ability to account for the diverse behaviours of agents and the underlying economic system makes ABMs a compelling alternative to conventional equilibrium-based economic models.

Gode and Sunder \cite{gode1993allocative} developed an agent-based model employing zero-intelligence traders to simulate financial markets. These traders lack strategic thinking, advanced learning capabilities, or statistical modelling of the financial market. \ce{However,}  despite their simplicity, zero-intelligence traders trade remarkably effectively, with prices converging to standard equilibrium levels and market efficiency reaching high levels. The study suggests that certain stylized facts in financial markets may depend more on institutional design than on actual agent behaviour. In other instances, agent-based models have been proposed to capture the "Trend" and "Value" effects in financial markets. For instance, Chiarella \cite{chiarella1992dynamics} designed an agent-based model featuring two types of traders: fundamentalists and chartists. Even with just these two trader types, the model exhibits numerous dynamic regimes that align with empirical evidence in artificial financial market simulations. An extension of the Chiarella model is proposed in \cite{MAJEWSKI2020103791}, introducing a new type of trader called noise trader and incorporating a long-term drift in the fundamental asset value. This extended model demonstrates more realistic price dynamics. 

The agent-based financial market simulation presented in this paper is built upon the Chiarella model from \cite{chiarella1992dynamics}. Instead of introducing noise traders in \cite{MAJEWSKI2020103791}, we introduce an innovative type of trader called volatility trader. The behaviour of the volatility trader is inspired by the Heston stochastic volatility model \cite{heston1993closed}. We present our proposed agent-based model for financial market simulation below.

\subsection{Deep reinforcement Learning and Deep Hedging}

% \ce{Compared with traditional RL,} one of the significant advantages of DRL is its ability to learn directly from high-dimensional input data, such as images or speech signals. Mnih et al. \cite{mnih2013playing} introduced \ce{Deep Q-Network (DQN)}, which demonstrated how deep neural networks could be used to learn control policies for playing Atari games directly from \ce{the game screen}. DQN was one of the first successful applications of DRL to high-dimensional visual inputs. Lillicrap et al. \cite{lillicrap2015continuous} introduced the Deep Deterministic Policy Gradient (DDPG) algorithm, extending DRL to continuous action spaces. DDPG has been widely used in robotics and control tasks. Heess et al. \cite{heess2015memory} modify the DDPG algorithm to use recurrent networks trained with backpropagation through time. They demonstrate that the resulting algorithms, Recurrent DPG (RDPG) can be applied to a number of partially observed physical control problems with diverse memory requirements.
% Later in \cite{mnih2015human}, they further improved the algorithm and achieved human-level performance on various Atari games, showcasing the potential of DRL for complex control tasks.

Deep Reinforcement Learning (DRL) is a type of machine learning that combines deep neural networks and reinforcement learning techniques to learn complex decision-making processes. DRL has been widely applied in various fields such as robotics, gaming, natural language processing, and computer vision. Popular DRL algorithms include Deep Q-Network (DQN) algorithm \cite{mnih2013playing} and Deep Deterministic Policy Gradient (DDPG) algorithm \cite{lillicrap2015continuous}. The idea of using deep reinforcement learning in finance is not new, but the application of these techniques in hedging has gained attention recently due to the need to hedge complex financial instruments, such as options on baskets of stocks or bonds, where traditional linear or non-linear models may not be effective. Bühler et al. \cite{deephedging} introduced the concept of deep hedging and demonstrated how deep learning techniques, such as neural networks, can be used to optimize hedging strategies for financial derivatives. The paper provides a theoretical framework for deep hedging and explores its applications in various financial settings. Horvath et al. \cite{risks9070138} trained the deep hedging agent with synthetic data generated from a rough volatility model, which provides a more accurate representation of market volatility dynamics compared to traditional models. The authors analysed the hedging performance of the original architecture under rough volatility models. Cao et al. \cite{cao2021deep} proposed a revised DDPG algorithm with two different Q-functions to derive optimal hedging strategies for derivatives when there are transaction costs. Their results showed that deep hedging can outperform traditional methods, such as delta hedging, in terms of hedging accuracy and computational efficiency.

% Deep hedging is a recent development that aims to create hedging strategies for complex financial instruments using deep reinforcement learning techniques. Later in \cite{buehler2020deep}, they extended the deep hedging framework to address the risk management problem of illiquid derivatives in a non-frictionless market, using a reinforcement learning approach.

We are aware of no existing work involving agent-based models in the deep hedging framework, which we aim to address in this paper. A brief review of the agent-based financial market simulation is presented above. In terms of the deep reinforcement learning method, we adopt the revised DDPG algorithm in \cite{cao2021deep}. The expected value of the hedging cost and the expected value of the square of the hedging cost are tracked by two different Q-functions. In this way, the standard deviation of the hedging cost can be included in the objective function. The learning objective of the deep hedging agent is to minimize a function equal to the mean hedging cost plus a constant times the standard deviation of the hedging cost.

\section{Proposed Chiarella-Heston Model}
In this paper, we propose an innovative agent-based model called Chiarella-Heston model. The model is inspired by both the extended Chiarella model in \cite{MAJEWSKI2020103791} and the Heston stochastic volatility model in \cite{heston1993closed}. It can also be considered a generalisation of both models. The model is proposed to address the shortcomings of the extended Chiarella model and the Heston model in generating realistic financial market simulations. For example, the extended Chiarella model is able to reproduce the "trend" and "value" effect in financial markets, but is difficult to replicate the volatility clustering stylised fact. On the other hand, the Heston volatility model is capable of reproducing the volatility clustering stylised fact, but it assumes a constant drift of asset price which is not the case in real financial markets. The proposed model overcomes these shortcomings by extending and combining the two models. Specifically, we replace the noise traders in the extended Chiarella model with volatility traders, whose demand level adapts according to the volatility signal level. From another perspective, We also replace the unrealistic constant drift in asset price in the Heston model with the demand from fundamental traders and momentum traders. In this way, a novel stochastic dynamic system is derived.

\subsection{Model Setup}
We denote the price of a stock at time $t$ as $P_t$. The total signed volume traded on the market from $t$ to $t+\Delta$ constitutes the cumulative demand imbalance in the same period. This quantity is denoted as  $D(t, t+\Delta)$. This aggregated demand depends on the trading strategies of various types of market participants. Following \cite{MAJEWSKI2020103791} and \cite{kyle1985continuous}, the price dynamics are assumed to be governed by a linear price impact mechanism:
\begin{equation} \label{kyleslambda}
P_{t+\Delta} - P_t = \lambda D(t, t+\Delta)
\end{equation}
where $\lambda$ is called "Kyle's lambda", which is related to the liquidity of the market and is a first-order approximation of market price sensitivity to market demand and supply. The market participants are assumed to be heterogeneous in their trading decisions. 

Following the Chiarella model \cite{chiarella1992dynamics}, we first populate our model with fundamental traders and momentum traders. Innovative volatility traders are then added to the model. Since traders of the same type exhibit the same behaviours, we only use one agent for each type of trader. This single agent represents the corresponding type of traders in the market and generates the aggregate demand and supply of that group of traders. With only three agents in the model, the simulation process is \ce{computationally efficient}. Each trader is associated with some parameters that control the trading behaviours and the amount of demand and supply generated by the corresponding trader group. We will show the calibration of these parameters in later sections.

\subsubsection{Fundamental Trader}
Fundamental traders base their trading decisions on the perceived fundamental value of a stock, which is denoted as $F_t$. If a stock's price is less than its fundamental value ($F_t - P_t > 0$), these traders are inclined to purchase it, and conversely, they are likely to sell it if it's overpriced. As per the standard set in \cite{chiarella1992dynamics}, it is assumed that the aggregated demand from fundamental traders is proportional to the degree of mispricing. Hence, the total demand from these fundamental traders is represented as $\kappa(F_t - P_t)$, with $\kappa$ controlling the overall demand generated by these traders. The fundamental value $F_t$ is an exogenous signal fed into the model.

\subsubsection{Momentum Trader}
Momentum traders, often referred to as "Chartists," engage in buying and selling financial assets based on recent price trends. Their strategy aims to capitalize on upward or downward movements in stock prices until the trend starts to fade. Unlike investors who consider the fundamental value of a stock, momentum traders primarily concentrate on analyzing recent price action and movement. They take long positions when the stock price has been on an upward trajectory and opt for short positions when the price shows a recent decline.

Numerous techniques exist for estimating the momentum of stock prices. One widely used approach is the exponentially weighted moving average of previous returns, where the decay rate is denoted as $\alpha$. As stated in \cite{MAJEWSKI2020103791}, this trend signal is represented as $M_t$:
\begin{equation} \label{momentumequation}
dM_{t} = -\alpha M_{t}dt + \alpha dP_t
\end{equation}
where $\alpha$ is the decay rate. Given the trend signal $M_t$, the demand function of momentum traders is represented as $f(M_t)$. According to \cite{MAJEWSKI2020103791}, the demand function $f(M_t)$ must satisfy two conditions:

\begin{itemize}
\item $f(M_t)$ is increasing.
\item $f''(M_t) * M_t < 0$
\end{itemize}
where the first condition aligns with the principles of momentum trading, while the second condition incorporates the risk-averse assumption applied to momentum traders. As described in \cite{chiarella1992dynamics}, we opt for the function $f(M_t) = \beta \tanh(\gamma M_t)$ with the constraint that $\gamma > 0$. In this context, $\gamma$ reflects the level of saturation in momentum traders' demand when momentum signals are exceptionally high. This saturation phenomenon can be attributed, in part, to factors like budget constraints and risk aversion, which are commonly observed among real chartists. Meanwhile, $\beta$ governs the overall demand generated by momentum traders and is assumed to be positive. This means that momentum traders exhibit positive demand when the momentum signal ($M_t$) is positive and negative demand when the momentum signal is negative. The selection of this demand function for momentum traders precisely fulfils the two specified requirements.

\subsubsection{Volatility trader}

We have incorporated a novel element into our agent-based model by introducing volatility traders. In addition to fundamental traders and momentum traders, a real financial market involves a diverse range of participants, each following distinct trading strategies. In our simplified model, volatility traders serve as a collective representation of all market participants influencing the dynamics of market volatility. Note that the trading activity of the volatility trader is different from the volatility trading in the derivative trading literature. In terms of trading logic, the volatility trader adjusts trading behaviour according to a volatility signal that mimics the real volatility dynamics. Heston model\cite{heston1993closed} is a popular stochastic volatility model used in finance to describe the dynamics of an asset's volatility over time. We choose the Heston model here because of nice features of the model such as the mean reversion of volatility and the correlation between an asset price and its volatility. Inspired by the Heston model, we introduce a volatility signal $\sqrt{\Sigma_t}$:
\begin{equation*}
d\Sigma_{t} = \phi (\theta - \Sigma_{t})dt + \sigma \sqrt{\Sigma_{t}} dW_{t}^{V}
\end{equation*}
where $\theta$ is the long-term mean of the variance of the asset returns, $\phi$ is the rate of mean reversion of the volatility process, and $\sigma$ is the volatility of the volatility process. These parameters can be estimated from empirical data, or be estimated globally to fit the stylised facts. The specific calibration process is presented in later sections. $W_t^{V}$ is a Wiener Process. $\sqrt{\Sigma_0}$ is the initial volatility level.

With the volatility signal $\sqrt{\Sigma_t}$ on hand, the demand function of the volatility trader is proportional to the volatility signal, with another Wiener Process included in the term: 
\[
% |D^{vol}_{t}| = \omega \sqrt{\Sigma_{t}}
D_{vol}(t, t+\Delta) = \omega \sqrt{\Sigma_{t}} dW_{t}^{S}
\]
where $\omega$ is a hyper-parameter that controls the overall demand level for volatility traders. $W_t^{S}$ is a Wiener Process that is correlated with $W_t^{V}$:
\[
% Corr(W_t^{S}, W_t^{V}) = \rho
dW_t^{S} dW_t^{V} = \rho dt
\]
where $\rho$ is the correlation coefficient. As is in the Heston model, the correlation here aims to create a correlation between the asset price and its volatility, which is a key feature of many financial markets and is known as the "leverage effect". The leverage effect refers to the observed phenomenon that asset returns and volatility are often negatively correlated. In other words, when the price of an asset decreases, its volatility tends to increase, and vice versa. By including a correlation between the demand from volatility traders and the volatility signal, the proposed model contributes to a correlation between the asset price and its volatility, making it a more realistic model of financial markets. The correlation is typically negative, reflecting the negative correlation between asset returns and volatility observed in many markets.

Overall, the demand function for the volatility trader is:

\begin{equation}
\begin{split}
&D_{vol}(t, t+\Delta) \hspace{0.2cm} = \omega \sqrt{\Sigma_{t}} dW_{t}^{S} \\
&d\Sigma_{t} \hspace{1.45cm} = \phi (\theta - \Sigma_{t})dt + \sigma \sqrt{\Sigma_{t}} dW_{t}^{V} \\
&dW_t^{S} dW_t^{V} \hspace{0.65cm} = \rho dt \\
\end{split}
\end{equation}

% \begin{equation}
%   \mathbb{\textbf{I}}=\begin{cases}
%     1,  \hspace{0.5cm} \text{if p <= 0.5}\\
%     -1, \hspace{0.3cm} \text{otherwise}
%   \end{cases}
% \end{equation}

\subsubsection{Model Dynamics}
Taking into account the demand of all three types of traders, the total demand is then given by
\begin{equation}
\begin{split}
D(t, t+\Delta) = \hat{\kappa} \int_{t}^{t+\Delta} (V_r - P_r)dr &+ \hat{\beta} \int_{t}^{t+\Delta} \tanh(\gamma M_r)dr \\
                 & + \hat{\omega} \int_{t}^{t+\Delta} \sqrt{\Sigma_r} dW_{r}^{S} .
\end{split}
\end{equation}
In addition, it is assumed that the fundamental value ($F$) is driven by a Geometric Brownian Motion with volatility $\sigma_F$ and drift $\mu$. We use $g$ to represent the volatility-adjusted average growth ($\mu - \frac{1}{2}\sigma_F^2$) at log level. Combined with Equation (\ref{kyleslambda}), the price dynamics for $\Delta \rightarrow 0$ is represented by the following stochastic dynamic system:

\begin{equation} \label{proposedmodel}
\begin{split}
dP_t &= \kappa(F_t - P_t)dt + \beta \tanh(\gamma M_t)dt + \omega \sqrt{\Sigma_t}dW_t^{S} \\
dF_t &= gdt + \sigma_F dW_t^{F} \\
dM_{t} &= -\alpha M_{t}dt + \alpha dP_t \\
d\Sigma_{t} &= \phi (\theta - \Sigma_{t})dt + \sigma \sqrt{\Sigma_{t}} dW_{t}^{V} \\
dW_t^{S} &dW_t^{V} = \rho dt \\
\end{split}
\end{equation}
where $\kappa$, $\beta$, $\omega$ are equal to $\lambda \hat{\kappa}$, $\lambda \hat{\beta}$, $\lambda \hat{\omega}$, respectively. Note that $P_t$ and $F_t$ are \textbf{logarithms} of prices and fundamental values, respectively.

\subsubsection{Discrete-time Simulation}
Equation (\ref{proposedmodel}) is the continuous version of our proposed model. To simulate and estimate the parameters of the model, a discrete-time version of the model is required. Here we present the discrete-time version of the proposed model, where one time step is equal to one day:

\begin{equation} \label{discretemodel}
\begin{split}
p_{t+1} - p_t &= \kappa (f_t - p_t) + \beta \tanh(\gamma m_t) + \omega \sqrt{\Sigma_{t}} \epsilon_{t+1}^{S} \\
m_{t+1} &= (1 - \alpha)m_t + \alpha (p_{t+1} - p_{t}) \\
f_{t+1} &= f_t + g + \eta_{t+1} \\
\Sigma_{t+1} &= \Sigma_{t} + \phi (\theta - \Sigma_{t}) + \sigma \sqrt{\Sigma_{t}} \epsilon_{t+1}^{V} \\
Corr(&\epsilon_{t+1}^{S}, \hspace{0.2cm} \epsilon_{t+1}^{V}) = \rho \\
\end{split}
\end{equation}
where $\eta_{t+1}$ is i.i.d. with normal distribution with zero mean and standard deviation $\sigma_F$. $\epsilon_{t+1}^{S}$ and $\epsilon_{t+1}^{V}$ both follow a normal distribution with zero mean and unit standard deviation. The correlation between $\epsilon_{t+1}^{S}$ and $\epsilon_{t+1}^{V}$ is $\rho$, which is generally negative. We remind again that $p_t$ and $f_t$ are logarithms of prices and fundamental values, respectively.

\subsection{Relationship with Existing Models}
Our proposed Chiarella-Heston model is a generalisation of the extended Chiarella model \cite{MAJEWSKI2020103791}. Observe that in Equation (\ref{proposedmodel}) for $\phi = 0$ and $\sigma = 0$ we recover the extended Chiarella model in \cite{MAJEWSKI2020103791}, with $\sigma_N = \omega \sqrt{\Sigma_0}$. The noise traders in the extended Chiarella model is described by Brownian Motion and have a constant parameter $\sigma_N$ describing the size of total demand from noise traders. Instead, the proposed Chiarella-Heston model replaces the noise traders with volatility traders, whose demand level adapts according to the volatility level. Intuitively, the volatility traders create larger demand (supply) level when the volatility signal is large, and vice versa. The volatility traders also contribute to reproducing the phenomenon that asset returns and volatility usually have negative correlation. Both features increase the realism of the proposed model and enable an advantage over the extended Chiarella model. In this sense, our proposed model can also be called an extended Chiarella Model with Stochastic Volatility.

On the other hand, the proposed Chiarella-Heston model can also be considered a generalisation of the Heston stochastic volatility model (\cite{heston1993closed}). Observe that in Equation (\ref{proposedmodel}) for $\kappa = 0$, $\alpha = 0$ and $\omega = 1$ we recover the Heston stochastic volatility model in \cite{heston1993closed}, with $\mu = \beta \tanh(\gamma M_0)$. The Heston model comprises one deterministic part and one non-deterministic part. The proposed Chiarella-Heston model converts the Heston model into an agent-based model by replacing the deterministic part. Specifically, the drift part in the Heston model is replaced by the aggregate demand from fundamental traders and momentum traders. The Heston model assumes a constant drift in asset price, while in reality, the asset price dynamic is not stationary. The proposed model drops the constant drift assumption. The interactions between fundamental traders and momentum traders contribute to the price dynamics in the model, making the model a more realistic replication of real financial market.

\subsection{Model Calibration} \label{modelcalibration}
This section outlines the methodology for calibrating the proposed Chiarella-Heston model. The calibration process involves identifying the best set of model parameters that allow the model to create the most accurate simulation of the financial market. Firstly, we discuss the actual data used and the corresponding stylized facts prevalent in financial markets. The workflow for calibrating parameters is then presented.

\subsubsection{Data and Stylised Facts}
During the model calibration process, acquiring real financial market data is imperative to establish an accurate calibration target. To achieve this, we collected daily price data for the S\&P 500 index spanning a total of 6000 days, encompassing the entire trading period from August 26th, 1999, to July 1st, 2023. The dataset is divided into two equal halves, each comprising 3000 consecutive days. The first 3000-day price data are employed for calibrating the model parameters, while the second 3000-day price data is reserved exclusively for testing purposes in the subsequent deep hedging experiments.

Financial price time series data display some interesting statistical characteristics that are commonly called stylised facts. According to \cite{sewell2011characterization}, stylized facts in financial markets refer to empirical findings that are so consistently present in asset returns across different markets and time periods that these findings are accepted as fundamental truths or "facts". A stylized fact is a simplified presentation of an empirical finding in financial markets. A successful and realistic financial market simulation is capable of reproducing various stylised facts. These stylised facts include fat-tailed distribution of returns, autocorrelation of returns, and volatility clustering. For a thorough review of financial stylised facts, refer to \cite{cont2001empirical}.

\subsubsection{Calibration Workflow}
\begin{figure}
  \includegraphics[width=0.5\textwidth]{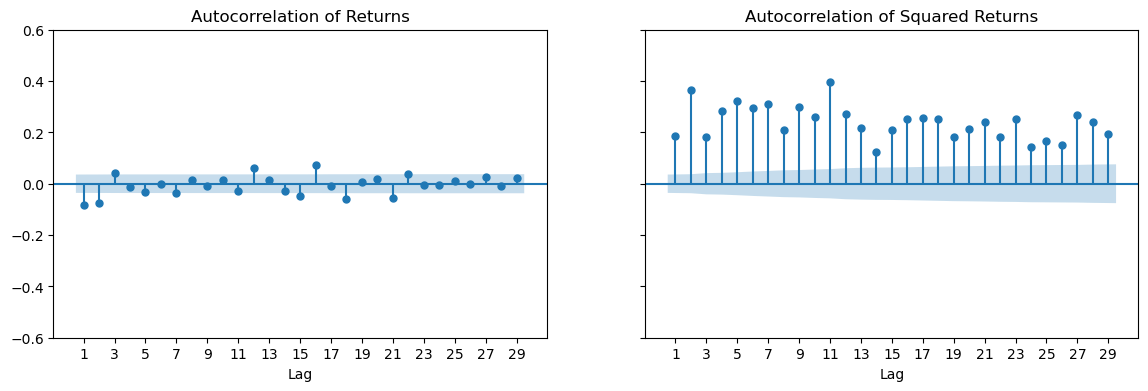}
  \caption{Historical stylised facts}
  \label{historical}
\end{figure}

In our calibration process, we mainly focus on three stylised facts: fat-tailed distribution of returns, autocorrelation of returns, and volatility clustering. The first one refers to the fact that price return distributions have been consistently observed to be fat-tailed across all timescales, indicating a higher likelihood of large price movements than what is predicted by standard normal distributions. The Hill Estimator of the tail index \cite{hill2010tail} is used as the primary metric for evaluating the fat-tail characteristic, with a lower value indicating a fatter tail in the return distribution. The second one refers to the fact that financial returns lack significant autocorrelation, except for small lags in high-frequency. Consequently, our daily simulation is required not to create significant autocorrelation patterns in returns. The third one, volatility clustering, refers to the phenomenon where periods of high volatility tend to be followed by periods of high volatility, and periods of low volatility tend to be followed by periods of low volatility. This stylised fact is reflected by the significantly positive autocorrelation of \textbf{squared} returns, slowly decaying with lags increasing. Figure~\ref{historical} presents the historical autocorrelation of returns and historical autocorrelation of squared returns.

The calibration target is to make the proposed model reproduce realistic stylised facts. A loss function is needed so that minimising the loss function improves the "realism" of a simulated financial market. Consistent with \cite{WILM:WILM11014} and \cite{gao2022high}, the loss function, called stylised facts distance, is constructed by measuring the discrepancy between the stylized facts from historical and simulated data. The stylised facts distance involves the above three stylised facts as well as the overall volatility level:
\begin{equation} \label{distancefunction}
D(\bm{\vartheta}) = w_1 * \Delta_{Hill}(\bm{\vartheta}) + w_2 * \Delta_{V}(\bm{\vartheta}) + w_3 * \Delta_{ACF^1}(\bm{\vartheta}) + w_4 * \Delta_{ACF^2}(\bm{\vartheta})
\end{equation}
where $\bm{\vartheta}$ is the model parameter vector to be estimated, $\Delta_{Hill}(\bm{\vartheta})$ is the distance between simulated Hill estimator and historical Hill estimator, $\Delta_{V}(\bm{\vartheta})$ is the difference between simulated volatility and historical volatility, $\Delta_{ACF^1}(\bm{\vartheta})$ is the distance between simulated autocorrelation of returns and historical autocorrelation of returns, $\Delta_{ACF^2}(\bm{\vartheta})$ is the distance between simulated autocorrelation of squared returns and historical autocorrelation of squared returns. $w_1$, $w_2$, $w_3$ and $w_4$ are corresponding weights. For detailed calculations of all terms in Equation~\ref{distancefunction}, refer to \cite{gao2022high}. Finally, the calibration target is:
\begin{equation} \label{thetahat}
\bm{\hat{\vartheta}} = \arg \; \min_{\bm{\vartheta} \in \bm{\Theta}} \; D(\bm{\vartheta}) \\
\end{equation}

\begin{figure}
  \includegraphics[width=0.5\textwidth]{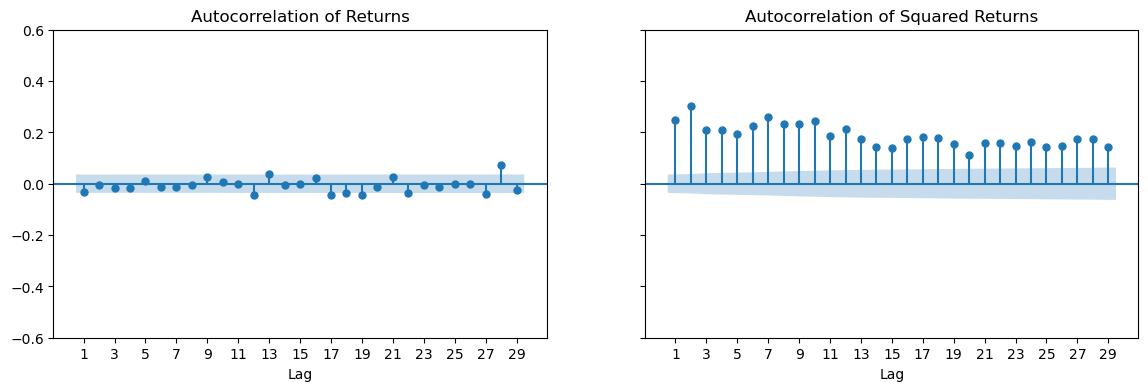}
  \caption{Chiarella-Heston model simulated stylised facts after calibration}
  \label{Chiarella}
\end{figure}

There are in total 11 model parameters to be estimated: $\kappa$, $\beta$, $\gamma$, $\omega$, $g$, $\sigma_F$, $\alpha$, $\phi$, $\theta$, $\sigma$ and $\rho$. Theoretically, we could estimate all 11 parameters by minimising the stylised facts distance. However, some of the parameters have economic representations and we could estimate these parameters directly from historical data. In this way, we also reduce computational costs during calibration. Following \cite{MAJEWSKI2020103791}, we fix parameter $\alpha$. We assume the typical horizon of trend computation for a low-frequency trend-following strategy is one week (five days), which gives $\alpha = 1 / (1 + 5) = 1 / 6$. We assume the fundamental value to be a Geometric Brownian Motion with drift ($\mu$) equals average historical return and standard deviation ($\sigma_F$) equals the average volatility of historical data. Recall that $g = \mu - \frac{1}{2}\sigma_F^2$. $\sigma$ takes the value of the volatility of the historical volatility, with window size 30 in our calculation. $\rho$ takes the value of the historical correlation between asset price returns and volatility. According to \cite{MAJEWSKI2020103791}, we also fix $\gamma$ and let $\gamma$ take the value of 10. The remaining model parameters, $\kappa$, $\beta$, $\omega$, $\theta$ and $\phi$ are calibrated via grid search by minimising $D(\vartheta)$. Figure~\ref{Chiarella} shows the simulated stylised facts after calibration, which resemble the historical stylised facts.

% $\theta$ is estimated by the long-term mean of the variance of the historical returns ($\theta = \sigma_F^2$).
% \begin{figure}
%   \includegraphics[width=0.5\textwidth]{gbm.png}
%   \caption{GBM stylised facts}
%   \label{GBM}
% \end{figure}

% The values for all model parameters after calibration are shown in Table~\ref{calibratedvalues}. 
We also calibrated the Geometric Brownian Motion, the Heston model, and the extended Chiarella model to match the stylised facts. The minimised stylised facts distances for all four models are shown in Table~\ref{stylisedfactsdistance}. It is illustrated that the proposed Chiarella-Heston model achieves minimum stylised facts distance among the four models. Specifically, the GBM and the extended Chiarella model are good at reproducing realistic volatility level, but are not able to capture the volatility clustering stylised fact (reflected by large $\Delta_{ACF^2})$. The Heston model could in a way replicate the volatility clustering stylised fact but are not good at reproducing realistic volatility levels. In contrast, the proposed Chiarella-Heston model can reproduce realistic volatility level as well as the volatility clustering stylised fact. In addition, the proposed model has another advantage over the other three models in that it replicates the fat-tails of returns in the financial market, shown by the much smaller $\Delta_{Hill}$. Overall, it is shown that the proposed model is a more realistic model than others.

% \begin{table}
%   \caption{Calibrated model parameter values}
%   \label{calibratedvalues}
%   \begin{tabular}{cccccccccccc}
%     \toprule
%     \makecell{Model \\ Parameter}&$\kappa$&$\beta$&$\gamma$&$\omega$ & $g$ & $\sigma_F$ & $\alpha$ & $\phi$ & $\theta$ &$\sigma$ & $\rho$\\
%     \midrule
%     \makecell{Parameter \\ Value} & 0 & 0 & 0 & 0 & 0 & 0 & $\frac{1}{6}$ & 0 & 0 & 0 & 0\\
%   \bottomrule
% \end{tabular}
% \end{table}

\begin{table}
  \caption{Comparison of stylised facts distance among different models after calibration}
  \label{stylisedfactsdistance}
  \begin{tabular}{ccccc}
    \toprule
    Model& GBM & Heston & \makecell{Extended \\ Chiarella} & \textbf{\makecell{Chiarella-\\Heston}} \\
    \midrule
    $\Delta_{V}$ & 0.002 & 0.229 & 0.003 & \textbf{0.028}\\
    \midrule
    $\Delta_{Hill}$ & 0.172 & 0.116& 0.177 &  \textbf{0.048}\\
    \midrule
    $\Delta_{ACF^1}$ & 0.067 & 0.064 & 0.069 & \textbf{0.066}\\
    \midrule
    $\Delta_{ACF^2}$ & 0.273 & 0.145 & 0.271 &  \textbf{0.082} \\
    \midrule
    \makecell{Total Stylised \\ Facts Distance} & 0.514 & 0.554 & 0.520 & \textbf{0.224}\\
  \bottomrule
\end{tabular}
\end{table}

\subsection{Model Validation}
To confirm the advantage of the Chiarella-Heston model over the baseline model, validation of the proposed model is carried out. Agent-based model validation in finance is the process of ensuring that an agent-based model accurately represents the real-world financial system. This involves comparing the outputs of the model under various scenarios to actual financial data. The main metric we use here is the Generalized Subtracted L-divergence (GSL-div), which is proposed in \cite{LAMPERTI201883}. 

We compare the GSL-div between the Chiarella-Heston model-generated time series and historical time series with the GSL-div between the GBM-generated time series and historical time series. Specifically, the calibrated Chiarella-Heston model is utilised to generate $M$ price scenarios. For each price scenario, a GSL-div value is calculated using the generated time series and historical time series. The same process is carried out for the GBM model. In this way, we obtain two sets of GSL-div values. A statistical test is carried out to compare the two sets. Results are shown in Table~\ref{statisticaltest}. The p-value is smaller than 0.05, showing that the GSL-div between the Chiarella-Heston model-generated time series and historical time series is significantly smaller than the GSL-div between the GBM-generated time series and historical time series. This result backs the conclusion that Chiarella-Heston model is able to generate more realistic simulated financial data.

\begin{table}
  \caption{Statistical test between two sets of GSL-div values}
  \label{statisticaltest}
  \begin{tabular}{cccc}
    \toprule
    \makecell{mean GSL-div \\ (Chiarella-Heston)}& \makecell{mean GSL-div \\ (GBM)} & t-statistic & p-value \\
    \midrule
    0.183 & 0.199 & 4.968 & 2.8e-6 \\
  \bottomrule
\end{tabular}
\end{table}

% \subsection{Risk Neutral Simulation}

\section{Deep Hedging Experiments}
Deep hedging is a technique utilising deep reinforcement learning to derive optimal hedging strategies for derivatives when there are transaction costs. Most existing methods apply Geometric Brownian Motion or the Heston model to generate simulated data to train and test deep hedging agents \cite{cao2021deep}. As an application of the proposed Chiarella-Heston model, we use the Chiarella-Heston model to generate simulated data for training a deep hedging agent. Instead of testing with simulated data, we test the performance of the deep hedging agent with real historical data. This same process is also carried out with the baseline Geometric Brownian Motion model, where training data are generated by a calibrated Geometric Brownain Motion. We show that the deep hedging agent trained by artificial data generated by Chiarella-Heston model achieves better performance than the agent trained by Geometric Brownian Motion data. 

\subsection{Problem Setting} \label{hedgeproblemsetting}
We consider a scenario where a trader is hedging a short position in a call option, where one option contract represents 100 underlying shares of the underlying asset. The trader hedges her option position solely by buying or selling the underlying asset. We make the assumption that the trader adjusts position at time intervals of 1 day and incurs trading costs. The option's lifespan is $N$ days, which is set to 30 in our experiments. In our model, the cost of a transaction in the underlying asset is proportional to the value of the asset being traded, although the model can be readily modified to suit different assumptions. Following \cite{cao2021deep}, the state at time $t$ comprises three values:
\begin{itemize}
    \item Current position of the underlying asset, which is determined by action from the previous time period ($t-1$ to $t$)
    \item Underlying asset price at time $t$
    \item Time to maturity at time $t$
\end{itemize}
At time $t= 0$ the initial underlying asset position is 0. The action at time t is to determine the amount of the underlying asset to be held for the next period; i.e., from time
$t$ to time $t+1$. The action space in our experiments is a continuous interval between 0 and 100. As for reward calculation, we opt for the accounting P\&L formulation, where the reward for each time step is the profit and loss of the combined portfolio including both the option position and the underlying asset position:
\begin{equation} \label{accounting}
R_{t+1} = -(V_{t+1} - V_t) + H_t(S_{t+1} - S_t) - \pi|S_{t+1}(H_{t+1} - H_t)| 
\end{equation}
for $0 \leq t \leq N$, where $V_t$ is the price of the option at time $t$, $S_t$ is the underlying asset price at time $t$, $H_t$ is the holding between time $t$ and $t+1$, $\pi$ represents the trading cost as a proportion of the transaction value. The $-1$ factor before $(V_{t+1} - V_t)$ in Equation~\ref{accounting} is because of the negative option position. The target of the problem is to train a deep reinforcement learning agent to drive an optimal trading strategy carried out in the span of $N$ days, which ideally minimises the accumulated hedging cost.
% , or in other words, maximises the accumulated profit and loss.

\subsubsection{Training and Testing Data}
We mentioned in Section~\ref{modelcalibration} that we collected daily price data for the S\&P 500 index spanning a total of 6000 days. The dataset is divided into two equal halves with the first half used for calibrating the proposed Chiarella-Heston model. The training data in our deep hedging experiments are generated by the calibrated Chiarella-Heston model. The calibrated Chiarella-Heston model is simulated according to Equations~(\ref{discretemodel}) for $M$ times ($M = 50000$ in our experiments), generating $M$ scenarios. Each scenario comprises a trajectory of the underlying asset price, with $N$ time steps. Option prices are calculated by the Black-Scholes model. The simulated training data constitute the training environment where the deep hedging agent learns to minimise the accumulated hedging cost.

The testing process utilises the second half of the collected S\&P 500 index daily price data to avoid any form of overlap between training data and testing data. The testing data consist of 3000 daily price observations. Using a rolling window with length $N = 30$, we obtain 2970 price trajectories. Calculating the returns of each price trajectory and providing an initial price, we obtain 2970 new price scenarios for testing the performance of the trained deep hedging agent. In this way, even though we change the scale of the price level, the testing price scenarios still have exactly the same returns as the historical price returns. The option price is again calculated by the Black-Scholes model. The justification for using the Black-Scholes model is that in reality, the prices of vanilla options are very close to the theoretical Black-Scholes price.

\subsection{Reinforcement Learning Algorithm}
The deep hedging agent in our experiments uses a revised form of the Deep Deterministic Policy Gradient (DDPG) algorithm \cite{lillicrap2015continuous} for learning the optimal hedging strategy. The algorithm here follows the set-up in \cite{cao2021deep} and works with hedging costs (negative rewards). The state space and action space are presented in Section~\ref{hedgeproblemsetting}. The objective function to minimise is the expected hedging cost plus a constant multiplied by the standard deviation of the hedging cost. As in \cite{cao2021deep}, two Q-functions are introduced to implement the objective function. One Q-function estimates the expected cost for state-action combinations while the other one estimates the expected value of the squared cost for state-action combinations. Since the detailed reinforcement learning algorithm is not the focus of this paper, please refer to \cite{lillicrap2015continuous} and \cite{cao2021deep} for detailed learning and updating rules of the DDPG algorithm applied in this problem setting.

The actor and critic neural network architectures in our experiments, though, are different from \cite{cao2021deep}. The actor and critic networks all have three hidden layers, with sizes 32, 64, and 32 neurons. The ReLU activation function is applied to each hidden layer. Batch normalization layers are also included between the input layer and the first hidden layer, as well as between adjacent hidden layers. For the actor-network, the output layer is followed by a Sigmoid function, mapping output to position levels. The training is carried out \ce{on} two NVIDIA GeForce RTX 3090 cards. Training for 15000 epochs costs approximately 5 hours.

\subsection{Deep Hedging Experimental Results}
\begin{figure}
  \includegraphics[width=0.45\textwidth, height=4.5cm]{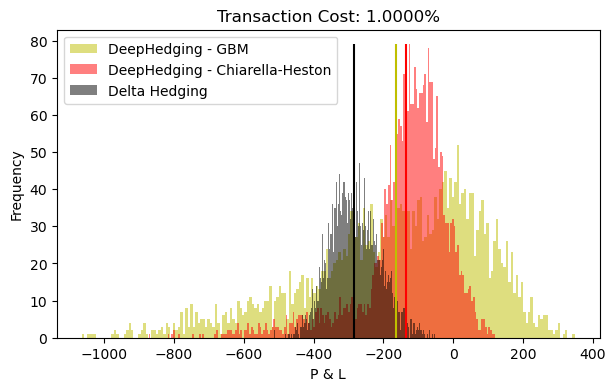}
  \caption{P\&L distribution of three hedging strategies}
  \label{hedgegraph}
\end{figure}

We train the deep hedging agent with the training dataset and test the performance of the learned hedging strategy on the testing dataset. The experiment is carried out under various transaction cost levels. In addition, we also obtain two baseline strategies for comparison. The first baseline is the traditional delta hedging strategy, which always keeps delta-neutral at the end of the day. The second baseline is the hedging strategy learned by the same deep hedging agent, but the training process is carried out on GBM-simulated data. The chosen performance metrics are the P\&L and the expected shortfall. Table~\ref{hedgepandl} and Table~\ref{hedgeexpectedshortfall} present the hedging performance (P\&L and expected shortfall) comparison between the deep hedging strategy trained on Chiarella-Heston model generated data and the two baselines - GBM-driven deep hedging strategy and traditional delta hedging strategy. The results show that the Chiarella-Heston model-driven deep hedging strategy clearly outperforms the other two strategies. In terms of P\&L, the two deep hedging strategies outperform the traditional delta hedging strategy in all transaction cost levels except 0.1\%. This is consistent with the literature that deep hedging generally performs better than delta hedging when transaction costs exist. According to Table~\ref{hedgepandl}, the two deep hedging strategies perform similarly in terms of P\&L. However in Table~\ref{hedgeexpectedshortfall}, the strategy trained by Chiarella-Heston model consistently performs better than the strategy trained by GBM data, achieving smaller expected shortfall. This shows that the Chiarella-Heston model successfully reproduces risky patterns in real financial markets, which is learned by the deep hedging agent to avoid losing too much in these scenarios. This successfully shows a practical application of the proposed Chiarella-Heston model. Figure~\ref{hedgegraph} shows a histogram of the P\&L distribution of the three strategies. It is shown very intuitively that the deep hedging strategy trained on data generated by Chiarella-Heston model significantly outperforms the other two strategies.

\begin{table}
  \setlength{\tabcolsep}{1.8pt}
  \caption{Hedging performance - P\&L (\$)}
  \label{hedgepandl}
  \begin{tabular}{ccccccc}
    \toprule
    Transaction Cost& 0.01\% & 0.1\% & 0.2\% & 0.4\% & 0.6\% & 1.0\% \\
    \midrule
    \makecell{Deep Hedging \\ Chiarella-Heston} & 54.45 & -43.68 & -22.62 & -80.64 & -185.26 & -136.37 \\
    \midrule
    \makecell{Deep Hedging \\ GBM} & 39.23 & -69.19 & 39.36 & -90.29 & -99.99 & -165.33 \\
    \midrule
    \makecell{Delta Hedging} & 8.62 & -17.93 & -47.44& -106.45 & -165.46 & -283.48 \\
  \bottomrule
\end{tabular}
\end{table}

\begin{table}
  \setlength{\tabcolsep}{1.8pt}
  \caption{Hedging performance - Expected Shortfall (\%)}
  \label{hedgeexpectedshortfall}
  \begin{tabular}{ccccccc}
    \toprule
    Transaction Cost& 0.01\% & 0.1\% & 0.2\% & 0.4\% & 0.6\% & 1.0\% \\
    \midrule
    \makecell{Deep Hedging \\ Chiarella-Heston} & -0.83\% & -3.33\% & -2.32\%& -2.94\% & -3.63\% & -3.36\% \\
    \midrule
    \makecell{Deep Hedging \\ GBM} & -2.08\% & -4.83\% & -2.47\% & -3.89\% & -4.81\% & -5.52\% \\
  \bottomrule
\end{tabular}
\end{table}

\section{Conclusions and Future Work}
\ce{The quantitative finance industry has seen a surge in data science applications, utilizing powerful algorithms with extensive data and computing capabilities. Deep learning algorithms are now being explored as replacements for traditional Greek-based methods in hedging derivatives, taking into account factors like transaction costs. Deep hedging, a data-driven approach that minimizes risk using neural networks, shows promise in optimizing hedge strategies. However, deep hedging demands substantial training data, often generated synthetically from stochastic models, which may not fully replicate real market dynamics. This research addresses this challenge by proposing an innovative Chiarella-Heston model, combining momentum, fundamental, and volatility traders. Calibrated to replicate empirical stylized facts, this model produces more realistic financial time series data, outperforming baseline mathematical models. The key contributions of this work are the Chiarella-Heston model and the enhanced performance of deep hedging agents in practical environments. The model's ability to generate accurate financial data and the trained agents' effectiveness in real trading situations highlight the potential of this methodology. By applying an agent-based approach to enhance deep hedging, this research provides a valuable strategy for improving risk management and hedging operations in the finance industry. The proposed model is validated using the Generalized Subtracted L-divergence metric. It generates a training dataset for a deep hedging agent, trained with the DDPG algorithm, to optimize hedging strategies under different transaction cost levels. The RL agent, trained with data from the proposed model, outperforms the baseline in most cost levels and performs well in real trading using empirical data.}

There are three directions of future work. Firstly, at the moment the proposed approach involves simulation of daily price data. For intra-day high-frequency simulation, it is worthwhile to develop a more advanced model architecture to replicate intra-day stylised facts. Secondly, other deep reinforcement learning algorithms may further boost the hedging strategy performance. Our future work would assess the performance of various deep reinforcement learning algorithms in deep hedging applications. Finally, we would also explore the possibility of deploying the proposed model for other practical applications. 

\bibliographystyle{ACM-Reference-Format}
\bibliography{sample-base}

\end{document}